\documentclass{article}
\usepackage{spconf,amsmath,graphicx}
\usepackage{multirow,url}
\usepackage{color,soul}
\usepackage{hyperref}
\usepackage{mathtools}
\usepackage{booktabs}

\title{Fast Conformer with Linearly Scalable Attention for Efficient Speech Recognition}
\name{\begin{tabular}{c}Dima Rekesh$^1$, Nithin Rao Koluguri$^1$, Samuel Kriman$^1$, Somshubra Majumdar$^1$,
Vahid Noroozi$^1$,
He Huang$^1$,\\
Oleksii Hrinchuk$^1$,
Krishna Puvvada$^1$,
Ankur Kumar$^2$,
Jagadeesh Balam$^1$,
Boris Ginsburg$^1$
\end{tabular}}

\address{$^{1}$NVIDIA, USA, \\$^{2}$Department of Computer Science, University of California, Los Angeles}
\copyrightnotice{979-8-3503-0689-7/23/\$31.00~\copyright2023 IEEE}
\begin{document}
%
\maketitle

\begin{abstract}

Conformer-based models have become the dominant end-to-end architecture for speech processing tasks. With the objective of enhancing the conformer architecture for efficient training and inference, we carefully redesigned Conformer with a novel downsampling schema. The proposed model, named \textit{Fast Conformer(FC)}, is 2.8$\times$ faster than the original Conformer, supports scaling to Billion parameters without any changes to the core architecture and also achieves state-of-the-art accuracy on Automatic Speech Recognition benchmarks. To enable transcription of long-form speech up to 11 hours, we replaced global attention with limited context attention post-training, while also improving accuracy through fine-tuning with the addition of a global token. Fast Conformer, when combined with a Transformer decoder also outperforms the original Conformer in accuracy and in speed for Speech Translation and Spoken Language Understanding. 

\end{abstract}
\noindent\textbf{Index Terms}: speech recognition, speech translation, spoken language understanding

\section{Introduction} 
\label{sec:intro}

\begin{figure}[t!]
    \centering    
    \includegraphics[width=1.0\linewidth]{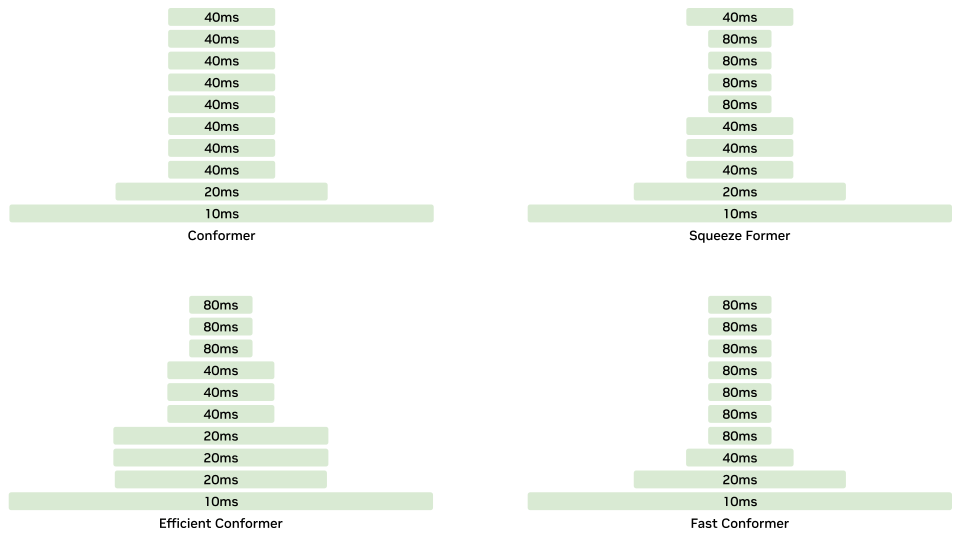}
    \caption{The downsampling schemas for Conformer, SqueezeFormer, Efficient Conformer and Fast Conformer.
    Fast Conformer increases sampling rate from 10ms to 80 ms using 3 depth-wise convolutional sub-sampling layers. The additional 2x reduction in the encoder output length versus  Conformer yields further compute-memory savings in RNNT decoder.
    }
    \label{fig:downsampling}
\end{figure}

Conformer is a  Transducer (RNNT) model  for automatic speech recognition (ASR) proposed by Gulati et al \cite{gulati2020conformer}. 
Conformer models obtain state-of-the-art results on multiple speech benchmarks\cite{guo2021espnet_conformer} thanks to  their encoder architecture which combines depth-wise convolutional layer for local features and self-attention layer for global context. Conformers have also been rapidly adopted in industry, especially  for streaming ASR on-device and in the cloud.
 \footnote{
\href{https://cloud.
google.com/blog/products/ai-machine-learning/
google-cloud-updates-speech-api-models-for-improved-accuracy}{F. Beaufays, “Google Cloud launches new models for more accurate Speech AI”, 2022}. 
\href{https://
cloud.google.com/blog/products/ai-machine-learning/
speech-on-device-run-server-quality-speech-ai-locally}{C. Barnes, “Run Google Cloud Speech AI locally, no internet connection required”, 2022}
}
At the same time, Conformer models use  more  compute and memory than convolution-only ASR models, e.g. Quartznet \cite{kriman2020quartznet}, because self-attention layers have quadratic time and memory complexity vs. input sequence length.
The quadratic complexity imposes a severe limitation on the maximum audio length which can be processed by Conformer. Scaling Conformer models require modifying conv kernel sizes\cite{zhang2020pushing} in Conformer blocks to stabilize large model training. 
We redesign Conformer to address these challenges.  Namely: 
\begin{enumerate}
  \item We redesign the downsampling schema and sub-sampling block to increase downsampling to 8x (see Fig.\ref{fig:downsampling})
  \item We optionally replace original self-attention layers with a combination of local attention and global context token post-training, similarly to LongFormer~\cite{longformer}.
\end{enumerate}
The proposed encoder has 2.9x fewer multiply-add operations with global attention, and can be made to scale linearly with sequence length post-training. It is  $2.8 \times$ faster than an equivalent Conformer during inference. It can scale to 1 Billion parameters without any changes to the core architecture. We hereby refer to this model as \textbf{Fast Conformer}.
At the same time, Fast Conformer maintains highly competitive word error rates (WER) on ASR benchmarks. 

We experiment with a Longformer-based attention mechanism, and find that by using limited context attention in combination with a single global attention token we are able to achieve good results on long-form ASR, while being more than $3 \times$ faster during inference. We then test Fast Conformer model on two additional speech processing tasks: speech translation (ST) and speech language understanding (SLU). On ST, with transformer and transduder decoders, we obtain strong scores for En-De translation with significant speedup compared to Conformer. On SLU we obtain state-of-the-art results on the Speech Intent Classification and Slot Filling task, and discuss the relatively modest speedup compared to the Conformer baseline. The models and training recipes have been open sourced through NVIDIA NeMo.\footnote{https://github.com/NVIDIA/NeMo}

\section{Fast Conformer architecture}
\subsection{Downsampling schema}

Conformer encoder is comprised of a stack of alternating multi-head attention \cite{attention2017}, depth-wise separable convolutional \cite{Chollet_2017} and fully connected layers with residual connections. The encoder starts with a sub-sampling module, which increases the frame rate from 10 ms to 40 ms. The 4x decrease in the sequence length helps reduce computational and memory costs of the attention layers in all following blocks.  
This subsampling module is relatively expensive, accounting for over 20 \% of the computation time for each forward pass of the model for the "Large" Conformer (120 M parameters) \cite{kim2022squeezeformer}. 

A straightforward way to  accelerate Conformer is to increase downsampling rate from 4x to 8x.
For example, \textit{EfficientConformer} \cite{burchi2021efficient_conformer} reduces the sequence length of the speech features by 8x using progressive downsampling: first,  2x down-sampling in the first layer, then another 2x in the middle of encoder, and final 2x in the last encoder layer. One of the drawbacks of  progressive subsampling  is a computation imbalance between different attention layers. The initial attention layers work on much longer sequences, so they have 16x more computational cost as compared to final attention layers that operate on much shorter sequences (see Fig. \ref{fig:downsampling}). 
Squeezeformer \cite{kim2022squeezeformer} combines  progressive downsampling with \textit{Temporal U-Net structure}.  Squeezeformer adds extra downsampling at the middle of encoder, and an upsampling layer at the end of encoder to recover 4x time resolution (see Fig. \ref{fig:downsampling}). A similar strategy was used in Uconv-Conformer~\cite{uconv_conformer}. 

One of the reasons why previous works  limit final encoder downsampling by 4x is related to the usage of the Conformer encoder with a CTC loss \cite{graves2006ctc}. CTC requires that the input to the loss function must be longer than the target sequence length.\footnote{Note that this constraint does not apply to the RNNT loss \cite{graves2012transducer}, and we are free to use any tokenization scheme as necessary. However, as the RNNT decoder is autoregressive, it is far more efficient at reducing the required number of calls to the Transducer Decoder and Joint by using large sub-word vocabularies.} 
Encoder output may become too short after 8x downsampling if the model uses character tokenization. For example, we found that  most of training samples in Librispeech (LS) \cite{panayotov2015librispeech} will not satisfy CTC condition after 8X subsampling if we use character tokenization.
To enable 8x downsampling for Conformer-CTC, we switch from character tokenization to Sentencepiece Byte Pair Encoding (BPE) \cite{kudo2018sentencepiece} with vocabulary sizes ranging from 128 to 1024 tokens. 

To accelerate Conformer, we made the following novel changes in the original design:
\begin{enumerate}
  \item 8x downsampling at the start of the encoder, thereby reducing the compute cost of subsequent attention layers by 4x 
  \item Replacement of the original convolution sub-sampling layers with depthwise separable convolutions \cite{chollet2017xception}
  \item Reduction of the number of convolutional filters in the downsampling block to 256
  \item Reduction of the convolutional kernel size to 9.
\end{enumerate}
A detailed comparison of Fast Conformer downsampling with previous works is presented in Table~\ref{tab:arch_differences}. 

\begin{table}[t]
\caption{Downsampling schemas and subsampling layer type for Conformer, EfficientConformer, Squeezeformer, and Fast Conformer. \textbf{K} is kernel size in convolutional filters. }
\label{tab:arch_differences}
\centering
\scalebox{0.85}
{
\begin{tabular}{l|llc}
 \toprule
   \textbf{Model} &
   \textbf{Subsampling schema} &
   \textbf{Type} &
   \textbf{K} \\
  \midrule
  {Conformer}\cite{guo2021espnet_conformer} &
    2/4     &
    2D Conv &
    31 \\
   {Squeezeformer}\cite{kim2022squeezeformer} &
    progressive 2/4/8/4  &
    Depth-wise sep  &
    31 \\
  {Eff. Conformer}\cite{burchi2021efficient_conformer} &
    progressive 2/4/8  &
    Depth-wise sep &
    15\\
   \midrule
  Fast Conformer & 
    2/4/8  &
    Depth-wise sep & 
     9  \\
  \bottomrule
\end{tabular}
}
\end{table}
To determine the contribution of each change to the model accuracy, we took Conformer-RNNT Large (115 M parameters) as a baseline and gradually applied each design  change. First, we added another 2x convolutional subsampling layer. Next, we used depthwise-separable convolutions in the second and third subsampling layers. Then we reduced the number of channels in the subsampling layers from 512 to 256. Finally, we reduced the convolutional kernel size in the conformer blocks from 31 to 9. 
Encoder inference speed was measured with a batch size of 128 on A100/80G GPU using 20s speech samples. Results are shown in Table \ref{tab:LibriSpeech}.
The encoder speed increased 2.8x, while maintaining model accuracy. 
\begin{table}
\centering
\caption{Accuracy vs speed for each component of Fast Conformer downsampling schema. Models were tested on LibriSpeech test-other incrementally for each modification starting from the original Conformer. 
Encoder inference speed (samples/sec) was measured with batch size 128 on 20 sec audio samples. The number of parameters (M) is shown for the encoder only.}
\label{tab:LibriSpeech}
\scalebox{0.8}{
\begin{tabular}{l|ccccc} 
 \toprule
\multirow{2}{*}{\textbf{Encoder}} & \textbf{WER, $\%$ } & \textbf{Inference,} & \textbf{Params,}& \textbf{GMACS}\\
      & \textbf{test-other} & \textbf{samples/s} & \textbf{M} \\
\midrule
Baseline Conformer     & 5.19 & 169 & 115 & 143.2\\
\  +8X Stride   & 5.11 & 303 & 115 & 92.5\\
\ \ +Depthwise conv     & 5.12 & 344 & 111 & 53.2\\
\ \ \ +256 channels & 5.09 & 397 & 109 & 48.8\\
\ \ \ \ +Kernel 9    & \textbf{4.99} & \textbf{467} & \textbf{109} & 48.7 \\
 \bottomrule
\end{tabular}
}
\end{table}

\subsection{Long-form audio transcription}

While standard multi-head attention layers have been successful in processing short utterances, their scalability to long sequences has been limited due to the quadratic scaling of the self-attention operation with sequence length. 
For example, Conformer can process at once  maximum 15 minutes audio on a single A100 GPU.  To address this challenge, several alternative approaches have been explored. A common approach is buffered transcription, where the audio sequence is divided into shorter chunks, which are then transcribed separately before being merged to form a complete transcription. 
Efficient-Conformer used \textit{grouped attention}  to reduce the cost of early attention layers   on long sequences by grouping neighboring time elements of the sequence along the feature dimension before applying scaled dot-product attention.

\begin{figure}[h]
    \centering    
    \includegraphics[width=0.9\linewidth]{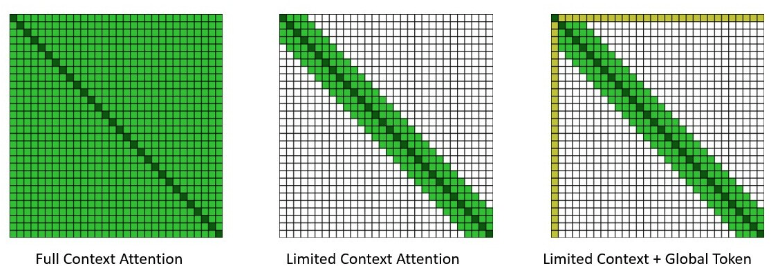} 
    \caption{Fast Conformer combines local attention with a global context token to process long speech samples. The figure is adapted from Longformer~\cite{longformer}}
    \label{fig:long_attention}
\end{figure}

We decided to use an alternative approach inspired by Longformer \cite{longformer} - local attention  augmented with global tokens. We use a single global attention token, which attends to all other tokens and has all other tokens attend to it, using a separate set of query, key and value linear projections, while others attend in a fixed-size window surrounding each token (see Fig. \ref{fig:long_attention}). 
By switching to limited context attention, we extend the maximum duration that the model can process at once on a single A100 GPU by 45x: from 15 minutes for unmodified Conformer to 675 minutes for Fast Conformer with limited context (see Table \ref{tab:max_dur}). In order to efficiently compute this attention, we utilize the overlapping chunks approach introduced in Longformer.

\begin{table}[t]
\caption{Maximum audio duration (min), which can be transcribed by model with batch size 1, A100 GPU.}
\label{tab:max_dur}
\centering
{
\begin{tabular}{l|c}
\toprule
\textbf{Model} & \textbf{Max duration, min} \\
\midrule
Conformer & 15    \\
Fast Conformer & 25  \\
Conformer + Limited Context & 135    \\
Fast Conf + Limited Context & 675  \\
\bottomrule
\end{tabular}
}
\end{table}

\section{Experiments}
\subsection{Automatic Speech Recognition}
The Fast Conformer model was evaluated on following English ASR benchmarks: LibriSpeech (LS)\cite{panayotov2015librispeech}, English part of Multilingual LibriSpeech (MLS)~\cite{MLS}., Mozilla Common Voice (MCV) \cite{Mozilla}, and Wall Street Journal (WSJ) \cite{wsj}. We used Fast Conformer-RNNT and Fast Conformer-CTC along with baseline Conformer models in \text{Large} configuration. 

\subsubsection{LibriSpeech}
First, we trained all models on the LibriSpeech dataset only.  We used the SentencePiece unigram tokenizer with 128 tokens for CTC, and 1024 tokens for RNNT models. Training of baseline Conformer-RNNT and -CTC  models, was done with AdamW optimizer with the Noam learning rate scheduler and peak learning rates of 0.0025 and 0.001 respectively. We set a linear warmup schedule for 15k steps in all experiments. Fast Conformer models were trained using cosine scheduler with peak learning rates of 0.005 and 0.001 respectively. Both Conformer and Fast Conformer-RNNT models were trained for 80k steps,  CTC models, for 380k steps. The models were trained on 32 GPUs with a global batch size of 2048. Last five checkpoints were averaged \cite{attention}. We used the Deepspeed profiler \footnote{\url{https://www.deepspeed.ai/tutorials/flops-profiler/}} to estimate Multiply Accumulate operations (MACS) on a single 30s audio. The results are shown in Table \ref{tab:LibriSpeech_bench}. Fast Conformer has slightly better accuracy than regular Conformer. The proposed encoder is 3x more compute efficient than original Conformer encoder, and is significantly faster than EfficientConformer and SqueezeFormer. 

\begin{table}[t]
\centering
\caption{ASR: Fast Conformer-Large with CTC and RNNT decoders trained on Librispeech. Greedy WER (\%) was measured on Librispeech test-other.  The number of parameters  (M) and compute (Multiply-Acc, GMAC) are shown for encoder only.}
\label{tab:LibriSpeech_bench}
{
\begin{tabular}{l|c|c|c} 
\toprule
\multirow{2}{*}{\textbf{Encoder}}    &
  {\textbf{WER, \%}}  &
  {\textbf{Params,}}   & 
  {\textbf{Compute,}}   \\   
  &
  \textbf{test-other} & 
  \textbf{M} & 
  \textbf{GMACS} \\
\midrule
\multicolumn{4}{c}{\textit{RNNT decoder}} \\
\midrule
Conformer  & 5.19 & 115 & 143.2 \\
\textbf{Fast Conformer} &\textbf{4.99} & \textbf{109} & \textbf{48.7}\\
\midrule
\multicolumn{4}{c}{\textit{CTC decoder}} \\
\midrule
Conformer &  5.74 & 121 & 149.2 \\
Eff. Conformer\cite{burchi2021efficient_conformer} &  5.79 & 125 & 101.3 \\
SqueezeFormer\cite{kim2022squeezeformer} &  6.05 & 125 & 91.0 \\
\textbf{Fast Conformer} & \textbf{5.64} & \textbf{115} & \textbf{51.5} \\
 \bottomrule
\end{tabular}
}
\end{table}

\begin{table}[t]
\centering
\caption{ASR: Fast Conformer-Large with CTC and RNNT decoders trained on English ASR set with 25K hours combined from public speech datasets. Greedy WER (\%) was measured on Librispeech test-other, MCV 8, MLS and WSJ-92 test sets }
\label{tab:asr3}
\scalebox{0.95}{
\begin{tabular}{l|cccc} 
 \toprule
\multirow{2}{*}{\textbf{Encoder}}    &
\textbf{LS } & \textbf{MCV 8} &
 \textbf{MLS}  & \textbf{WSJ-92} \\
  & \textbf{test-other} & \textbf{test} & \textbf{En} & \textbf{test}  \\
\midrule
\multicolumn{5}{c}{\textit{RNNT decoder}} \\
\midrule
\textbf{Conformer}  & \textbf{3.74} & \textbf{7.87} & 5.77 & 1.47  \\
Fast Conformer & 3.79 & 8.18 & \textbf{5.76} & \textbf{1.42}  \\
\midrule
\multicolumn{5}{c}{\textit{CTC decoder}} \\
\midrule
Conformer  & 4.50 & 9.40 & 6.60 & 1.70  \\
\textbf{Fast Conformer} & \textbf{4.19} & \textbf{9.00} & \textbf{6.42} & \textbf{1.59} \\
 \bottomrule
\end{tabular}
}
\end{table}

\subsubsection{Large-25k hours NeMo ASR Set}
To test Fast Conformer capacity with respect to larger dataset, we trained Fast Conformer and Conformer models on 25k hours  of speech, composed from LibriSpeech, Mozilla Common Voice, National Singapore Corpus and other public English speech datasets. 
All RNNT models were trained for 300k steps and CTC models for 1M steps. AdamW was used with cosine scheduler with a 15k linear warmup and learning rates of 0.0025 and 0.001, for RNNT and CTC respectively. 
The models have been tested on LibriSpeech test-other, MLS, MCV and WSJ-93 test sets. The results are presented in Table \ref{tab:asr3}. Fast Conformer outperformed the original Conformer on most benchmarks.

\begin{table}[t]
 \caption{Speech Translation, MUST-C V2 tst-COMMON dataset. SacreBLEU, total inference time, and relative inference speed-up were measured with a batch size $32$ for two speech translation models with Conformer-based encoder and either RNNT, or Transformer decoder.}
 \centering
 \label{tab:st_results}

  \begin{tabular}{lccc}
  \toprule
  \textbf{Encoder} & \textbf{BLEU} & \textbf{Time, sec} & \textbf{Speed-up} \\
 \midrule
\multicolumn{4}{c}{\textit{Transformer decoder}} \\
\midrule
  Conformer  & 31.0 & 267  &  1X \\
    Fast Conformer & 31.4 & 161 & 1.66X \\  
  \midrule
\multicolumn{4}{c}{\textit{RNNT decoder}} \\
\midrule
  Conformer & 23.2 & 83 & 1X \\
  Fast Conformer & 27.9 & 45 & 1.84X \\
  \bottomrule
  \end{tabular}

\end{table}

\subsection{Speech Translation}
Next, we analyze the efficacy of Fast Conformer on Speech Translation (ST) from English to German. 
We trained two architectures with the same Conformer-like encoder and different autoregressive decoders: either RNNT, or 6-layer Transformer trained with cross entropy loss.

In all the experiments, we initialized encoder with the corresponding weights from ASR RNNT models trained on 25k hours of speech. The parameters of decoder and joint module were initialized randomly. Our vocabulary consists of $16384$ YouTokenToMe\footnote{\url{https://github.com/VKCOM/YouTokenToMe}} byte-pair-encodings trained on German text. Our models have been trained on all available datasets from IWSLT22~\cite{anastasopoulos2022findings} competition which corresponds to ~$4$k hours of speech. Some of the datasets did not contain German translations, so we generated them ourselves with text-to-text machine translation model trained on WMT21~\cite{subramanian2021nvidia} and in-domain finetuned on Must-C v2\cite{cattoni2021must}.

The results for all models are shown in Table \ref{tab:st_results}. We note that RNNT loss is generally not suitable for speech translation due to its implicit monotonic alignment assumption. Surprisingly, Fast Conformer-RNNT translation model gets BLEU score of 27.89. In addition, the inference of this model is up to 1.84$\times$ faster compared to Conformer. 

\subsection{Spoken Language Understanding}

\begin{table}[t]
\caption{Speech intent classification and slot filling on SLURP dataset. ESPNet-SLU  and SpeechBrain-SLU models use a HuBERT~\cite{hsu2021hubert} encoder pre-trained via a self-supervised objective on LibriLight-60k~\cite{kahn2020librilight}. 
Inference time and relative speed-up against Conformer are measured with batch size 32.}
\label{tab:slu}
 \resizebox{1.0\columnwidth}{!}
{
\begin{tabular}{l|cccc}
\toprule
\textbf{Model} & 
\begin{tabular}[c]{@{}c@{}} 
\textbf{Intent} \\ \textbf{Acc.} 
\end{tabular} & 
\begin{tabular}[c]{@{}c@{}}\textbf{SLURP}  \\ \textbf{F1} \end{tabular} & 
\begin{tabular}[c]{@{}c@{}}\textbf{Inference,}\\ \textbf{sec} \end{tabular} &
\begin{tabular}[c]{@{}c@{}}\textbf{Rel.} \\ \textbf{Speed-up} \end{tabular} \\ 
\hline
SpeechBrain-SLU & 87.70 & 76.19 & - & - \\
ESPnet-SLU      & 86.52 & 76.91 & - & - \\
\midrule
\multicolumn{5}{l}{\textbf{Conformer/Fast Conformer+Transformer Decoder}} \\
\midrule
Conformer & 90.14 & 82.27 & 210 & 1X \\
\textbf{Fast Conformer} & \textbf{90.68} & \textbf{82.04} & \textbf{191}& \textbf{1.1X} \\
\bottomrule
\end{tabular}
}
\end{table}

Next, we apply the pre-trained Fast Conformer encoder to spoken language understanding (SLU). We study the \emph{Speech Intent Classification and Slot Filling} (SICSF) task, which should detect user intents and extract the corresponding lexical fillers for detected entity slots ~\cite{slurp2020}. An intent can be a composition of a scenario type and an action type. Slots and fillers are represented by key-value pairs. The ground-truth intents and slots of input are organized as a Python dictionary, represented as a string. The SICSF task is to predict this structured Python dictionary as a string, based on the input audio. Experiments are conducted using the SLURP~\cite{slurp2020} dataset, where intent accuracy and SLURP-F1 are used as the evaluation metric. 

We use as the baseline Conformer encoder, initialized from pretrained ASR model, with a Transformer decoder.
We also compare Fast Conformer against two state-of-the-art models from ESPNet-SLU~\cite{arora2022espnet} and SpeechBrain~\cite{wang2021fine}. ESPNet-SLU  and SpeechBrain both use a HuBERT~\cite{hsu2021hubert} encoder pre-trained via a self-supervised objective on the entire LibriLight-60k~\cite{kahn2020librilight} dataset. ESPNet-SLU further finetunes the encoder on LibriSpeech before training on SLURP. 

The results for SLURP experiments are shown in Table~\ref{tab:slu}. The model with pre-trained Fast Conformer encoder  significantly surpass the ESPNet-SLU and SpeechBrain  models, which have been pre-trained on nearly 60K hours of speech. Fast Conformer attains very high accuracy quite close to Conformer.  Its decoding  is also 10(\%) faster than Conformer. The speed-up is not as high as for ASR for following reason:  the ratio of acoustic signal length (after 8x downsampling) to target token length is roughly \textbf{1:2.22} for the SICSF task. The  execution cost for encoder is dwarfed by slow autoregressive Transformer decoder, and therefore we used  batch 32 to balance the cost of the encoder against the decoder to show speed-up.

\begin{table}[t]
\caption{Fast Conformer versus Conformer on long audio. We evaluated four versions of Fast Conformer: (1) FC with full context attention (2) FC with limited context (3) FC with limited context and global token. Models have been evaluated on two long-audio bencmarks: TED-LIUM v3 and Earnings 21. Normalized greedy WER(\%).}
\label{tab:long_asr}
\begin{tabular}{l|c|c}
\toprule
\textbf{Model} & \textbf{TED-LIUM v3} & \textbf{Earnings21}  \\
\midrule
Conformer & 9.18 & 18.26 \\
\midrule
Fast Conformer (buffered) & 9.15 & 17.65 \\
\ + Limited Context & 8.25 & 16.08 \\
\ \ + Global Token & 7.5 & 11.85 \\
\bottomrule
\end{tabular}
\end{table}
\subsection{Long-form audio transcription}

Fast Conformer with limited context and global token for long audio was trained   in the following way. This model has shared query, key and value projection layers that are used for global and local attention.
The encoder is initialized with the checkpoint pre-trained on our internal 25k hour set with full context.  We then do additional fine-tuning on the same dataset with limited context attention for 10k steps. We used learning rate  warmup of 1k steps, with maximum learning rate 1e-6 and cosine rate annealing to 0. The size of the limited context was set to 128 steps on each side of a token, which corresponds to around 10 seconds.

The performance of Fast Conformer  was evaluated using two long-form audio datasets: TED-LIUM v3~\cite{hernandez2018ted}, and Earnings-21~\cite{earnings21}. We compare our limited context model with full context Fast Conformer as well as with base Conformer trained on the same dataset. We used the Whisper normalizer on both transcripts and predictions to evaluate the models.
For Conformer and Fast Conformer with full context attention we used 20 second buffers.
For Fast Conformer with limited context we processed the full audio in one forward pass.  Fast Conformer with new attention mechanism significantly outperforms Conformer and Fast Conformer with global attention on both long-form ASR benchmark sets  (see Table \ref{tab:long_asr}).

\section{Scaling Fast Conformer Model}
\label{sec: scaling}

\begin{table}[t]
\caption{Table presents parameter modifications required for constructing the FC-L, -XL, and -XXL models. Increased the hidden dimension (d\_model), encoder layers and decoder RNNT layers to build XL from L. Keeping other model parameters constant from XL, we increased the number of encoder layers to construct FC-XXL.}
\centering
\begin{tabular}{c|c|c|c|c}
\toprule
\textbf{Model}                     & \textbf{\begin{tabular}[c]{@{}c@{}}Hidden\\    Dimension\end{tabular}} & \textbf{\begin{tabular}[c]{@{}c@{}}Encoder\\    Layers\end{tabular}} & \textbf{\begin{tabular}[c]{@{}c@{}}RNNT \\ Layers\end{tabular}} & \textbf{\begin{tabular}[c]{@{}c@{}}Model\\    Parameters\end{tabular}}\\ \midrule
L & 512 & 17 & 1 & 120 M \\
XL & 1024 & 24 & 2 & 600 M \\
XXL & 1024 & 42 & 2 & 1.1 B \\ \bottomrule
\end{tabular}
\label{tab:model_sizes}
\end{table}

To show scaling capacity of Fast Conformer models, we designed three model sizes: Large (L), Extra Large (XL) and Extra Extra Large (XXL), similar to Conformer scaling in \cite{zhang2020pushing} as shown in Table: \ref{tab:model_sizes}. 

 We observed that when scaling FC models from XL to XXL, pretraining the encoder with Self Supervised learning helps stabilize training and enable high learning rates.  We adopted the pretraining and finetuning method of SSL models based on 
Wav2Vec 2.0 \cite{baevski2020wav2vec}. Unlike Conformer models\cite{zhang2020pushing}, we didn't change the conformer blocks and relative attention while scaling up models. From -L to -XXL core architecture of all FC models remains the same. 

\begin{table}[htbp]
    \caption{Comparison of XL and XXL models on ASR benchmark datasets. Table illustrates the performance comparison between Conformer-XL versus Fast Conformer-XL RNNT models, as well as the improvement of Fast Conformer-XXL over Fast Conformer-XL. All models are trained on 25k hrs of ASR Set. Greedy WER(\%).}
    \centering
     \resizebox{1.0\columnwidth}{!}
{
    \begin{tabular}{c|c|c|c|c}
        \toprule
        \textbf{Model} & $\begin{array}{c}\text {\textbf{LS}} \\
        \text {\textbf{Test-clean}}\end{array}$ & $\begin{array}{c}\text {\textbf{LS}} \\
        \text {\textbf{Test-other}}\end{array}$ & $\begin{array}{c}\text {\textbf{MLS}} \\
        \text {\textbf{Test}}\end{array}$ & \textbf{GMACS}\\
        \midrule
         Conformer-XL & 1.49 & 2.80 & 5.32 & 686\\
         \midrule
         FC-XL & 1.50 & 2.88 & 4.90  & 253\\
         FC-XXL & 1.38 & 2.52 & 4.58 & 441 \\
        \bottomrule
    \end{tabular}
    \label{tab:scale_fc}
}
\end{table}

\begin{table*}[ht!]
\caption{The performance of FC-XL and FC-XXL models was compared with an augmented training set. The table illustrates the performance improvement of both models as we augment the training dataset by an additional 40,000 hours (ASR Set ++) on HF-audio leaderboard test sets~\cite{open-asr-leaderboard} after whisper text normalization~\cite{whisper} applied. Greedy WER (\%).}

 \centering
 \scalebox{0.86}{
    \begin{tabular}{c|c|c|c|c|c|c|c|c|c|c|c}
        \toprule 
        \multirow{2}{*}{\textbf{Model}} & \multirow{2}{*}{\textbf{Decoder}} & \textbf{Train} & \textbf{LS}  & \textbf{LS} & \textbf{TED-LIUM} & \textbf{Vox} & \textbf{MCV 9} & \textbf{AMI} & \textbf{Earnings} & \textbf{SPGI} & \textbf{Giga} \\
        & & \textbf{Dataset} & \textbf{Test-clean} & \textbf{Test-other} & \textbf{V3} & \textbf{Populi} & \textbf{Test} & \textbf{Test} & \textbf{22} & \textbf{Speech} & \textbf{Speech}\\
        \midrule 
        \multirow{4}{*}{FC-XL} & \multirow{2}{*}{RNNT} & ASR Set & 1.50 & 2.88& 4.49& 5.74& 7.26& 18.28& 16.37& 4.40 &  11.58\\
        & & ASR Set ++ & 1.63& 3.06& 3.86& 6.05& 8.07& 17.55& 14.78& 3.47& 10.07\\
        \cmidrule{2-12}
        & \multirow{2}{*}{CTC} & ASR Set & 1.73& 3.47& 4.71& 6.09& 7.51& 18.41& 17.89& 5.04&11.84\\
        & & ASR Set ++ & 1.87& 3.76& 3.78& 7.00 & 10.57&16.3& 14.14& 4.11&10.35\\
        \midrule
        \multirow{4}{*}{FC-XXL} & \multirow{2}{*}{RNNT} & ASR Set & \textbf{1.38}& 2.52& 4.74& 5.56& 6.07& 18.81& 16.66& 4.98&11.95\\
        & & ASR Set ++ & 1.46& \textbf{2.47}& 3.92& \textbf{5.39}& \textbf{5.79}&17.1& 14.11& \textbf{3.11}&\textbf{9.96}\\
        \cmidrule{2-12}
        & \multirow{2}{*}{CTC} & ASR Set & 1.69& 3.4& 4.64& 6.45& 8.31&17.62& 16.44& 4.91&11.61\\
        & & ASR Set ++ & 1.83& 3.54& \textbf{3.54} & 6.53& 9.02&\textbf{15.62}& \textbf{13.69}& 4.20 &10.27\\
        \bottomrule
    \end{tabular}
    }
    \label{tab:large-dataset}
\end{table*}

    The XL and XXL models yield superior results within fewer training steps when compared to L models. As shown in Table \ref{tab:scale_fc} when evaluated on HF-Leaderboard\cite{open-asr-leaderboard} evaluation test sets \footnote{\url{https://huggingface.co/spaces/hf-audio/open_asr_leaderboard}}, performance analysis showcases the effectiveness of the FC-XL and FC-XXL models trained on 25k hrs of NeMo ASR Set. The -XL models presented were trained for 70k steps, while -XXL the models were trained for 100k steps with an effective batch size of 2048. For the XL model, we use Adamw optimizer with Noam learning rate scheduler\cite{vaswani2017attention} with peak learning rate of 6e-4 and 15k linear warmup steps, whereas FC-XXL models were linearly warmed up for 25k steps. All XL and XXL models were initialized with pretrained SSL checkpoints. We also observed that finetuning a RNNT FC-XL model with CTC just for 40k steps showed similar performance to training FC-XL CTC model for 200k steps from scratch. 

\subsection{Scaling Dataset}

\begin{figure}[htpb]
    \centering
    \includegraphics[width=0.9\linewidth]{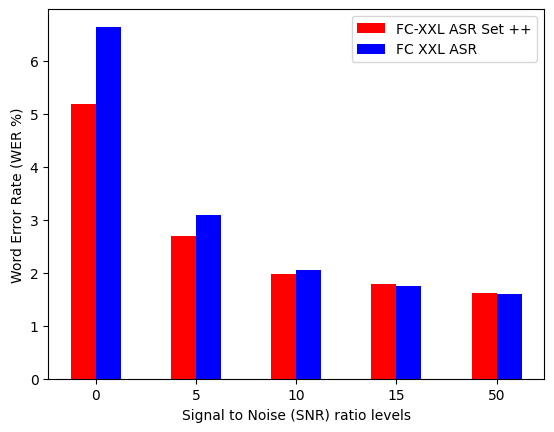}
    \caption{Figure demonstrates the noise robustness of FC-XXL models on the LS Clean evaluation set, showcasing their performance across various signal-to-noise ratio (SNR) levels.}
    \label{fig:SNR-robustness}
\end{figure}

The FC-XXL RNNT model, which was trained on 25k hours of ASR Set, achieves similar state-of-the-art performance on LS-test other as \cite{zhang2020pushing}, while also achieving the best performance on other benchmark datasets. However, in order to effectively leverage large models, it is imperative to proportionally augment dataset sizes in accordance with the model sizes. Hence, we trained these large models by incorporating an additional 40,000 hours of internal dataset (ASR Set ++). The integration of these supplementary datasets facilitated enhanced accuracy and noise robustness in both XL and XXL models. 
Table \ref{tab:large-dataset} shows successful utilization of large amount of data by effectively training the 1B parameter model. Furthermore, Figure \ref{fig:SNR-robustness} illustrates the noise robustness of the FC-XXL models across different signal-to-noise ratio (SNR) levels on the clean set of the Librispeech dataset.

\section{Conclusions}
\label{sec:conclusions}

In this paper, we propose Fast Conformer, a redesigned Conformer with a novel downsampling schema. Fast Conformer uses 2.9x less compute while delivering roughly the same WER as the original Conformer. Evaluations on tasks such as speech translation and spoken language understanding show a strong model accuracy while achieving significant speed-ups in the encoder computation. When the attention module is replaced with local attention, we show that the greater efficiency enables long-form transcription of an 11-hour audio segment in a single forward pass. The results on long-form audio are improved further by adding a single global attention token. We finally show that Fast Conformer architecture can be easily scaled to 1B parameters which enables us to further improve accuracy while achieving noise robustness when training on larger datasets.


\bibliography{refs_v2}
\bibliographystyle{IEEEbib}

\end{document}